\documentstyle[11pt,newpasp,twoside,epsf]{article}
\def\edcomment#1{\iffalse\marginpar{\raggedright\sl#1\/}\else\relax\fi}
\reversemarginpar

\begin{document}
\title{ Kinematics of the Galactic populations in the GAIA era}
 \author{G. Bertelli$^1$, A. Vallenari$^1$, S.Pasetto$^{1,2}$, C. Chiosi$^1$}
\affil{$^1$ INAF, Osservatorio di Padova, Vicolo Osservatorio 5, 35122 Padova}
\affil{$^2$ INAF, Dipartimento di Astronomia, Vicolo Osservatorio 3, 35122 Padova}

\begin{abstract}

GAIA data will create a precise 3-dimensional map of  the Galaxy, providing positional information, radial
velocities, luminosity, temperature and chemical composition of
 a representative
sample of stars.
Here
we present a new implementation of the Padova Galaxy Model,
where kinematics simulations are included.
A few examples of application to
 GAIA science are discussed, in particular concerning radial velocities
determinations of the thin and thick disk. 
\end{abstract}

\section{Introduction}
GAIA  will 
provide detailed phase space coordinates for about 1 billion stars within a sphere of 20 Kpc.
In addition to this  on board multicolour photometry and spectroscopy will give
complete chemical measurements (including $\alpha$, $s$, $r$ process elements)
down to V=17-19 mag.
Owing to the precision of the data, the problem
of the formation of the Milky Way will be addressed.
 In the standard cold dark matter scenario the Galaxy should have formed by  merges of small substructures. 
However this scenario is at odds with several observational constraints:
i.e. a  small scale length of the disk of the order of 300 pc is expected
 instead of 2000-3000 pc  (Steinmetz \& Navarro 1999); 
the formation process predicts that the Milky Way should have a
thousand of satellites which are missing in the Local Group.
 A substantial revision of the model
is necessary, by means of a detailed comparison with high quality data.
GAIA will bring into evidence fossil remnants of the
galaxy formation tracing back the star formation history and revealing gradients
in the star formation intensity, chemical abundances, and kinematics across the disk.
Chemical and age gradients are not expected inside the thin disk.
  N-body simulations show that radial mixing can wash out gradients
close to the Galactic plane after a few Gyr.
Concerning the thick disk, if it was formed by an heating event (merge with a satellite) traces of it should still be detectable (Freeman \& Hawthorn 2002).
In fact vertical gradients in  metallicity and velocity dispersions 
which are still uncertain from the present data  are possibly not
 washed out by the
radial mixing far from the Galactic plane and can be used to trace 
back the formation history. In order to simulate Galactic data 
 we update the Padova Galaxy Model including velocity  space and we apply it to the GAIA science.

 \section {Padova Galaxy Model}

 The Galaxy is modeled with the code already described  by Bertelli 
et al. (1995) and revised as  in Vallenari et al (2000) where more detail can be found.
The Padova model  has been newly updated including:

1) the usage of new stellar tracks from Z=0.0001 till Z=0.03 
with low mass stars down to 0.15 
(Girardi et al 2000). 0.1 M$_\odot$ track is taken from  Baraffe et al
(1998);

2) the use  of the
 carbon star models taking into account the effect of the variation of the
stellar molecular  opacities
during the evolution as a result of the dredge-up (Marigo 2002);

3) the extinction along the line of sight
 derived following   Drimmel \& Spergel (2001) model obtained from COBE-DIRBE
infrared data.  

4) the velocity space  simulated in a consistent way as described in the following Section.

\section {The kinematic model}

The velocity distribution for the whole disk has been computed using the velocity ellipsoids
formulation by Schwarzschild. Concerning the thin disk,
the assumed local values of velocity dispersions  are taken as in Mendez
et al (2000).
The V$_{lag}$ is derived following Binney \& Tremaine (1994). The 
diagonal terms of the dispersion velocity tensor are from
 Lewis \& Freeman (1989):

\begin{eqnarray}
\sigma^2_{RR} &=& \sigma^2_{RR,0}exp(-(R-R_0)/H_R)\\
\sigma^2_{\Phi \Phi} &=& 1/2 (1+(dln V_{LSR} (R))/(dlnR)) \times \sigma^2_{RR})\\
\sigma^2_{ZZ}&=&\sigma^2_{ZZ,0}exp(-(R-R_0)/H_{R})
\end{eqnarray}

The vertical gradient in the velocity dispersions is taken from Fuchs \& Wielen (1987)
for $\sigma^2_{RR}$ and $\sigma^2_{\Phi \Phi}$ and the non vertical isothermality of the thin 
disk from Amendt \& Cutterford (1991) which is in good agreement with the observations till
 1 Kpc
for $\sigma^2_{ZZ}$. The off diagonal term  $\sigma_{RZ}$ is derived from Amend \& Cutterford 
(1991):

\begin{equation}
\sigma_{RZ}^2(R,Z)=\sigma^2_{RZ}(R,0)+z\partial/{\partial z} \sigma^2_{R,Z}(R,0)
\end{equation}

where 

\begin{equation}
\partial/{\partial z} \sigma^2_{RZ}(R,0)= \lambda(R)({\sigma^2_{RR}-\sigma^2_{ZZ}}/R)(R,0)
\end{equation}

and

\begin{equation}
\lambda(R)=(R^2 \Phi_{Rzz}/{3\Phi_{R}+R \Phi_{RR}-4R\Phi_{ZZ}})(R,0) 
\end{equation}

where $\Phi_{RZZ}, \Phi_{RR},\Phi_{ZZ}$ are the derivatives of the
Galactic potential 
$\Phi$  obtained from the density model by Dehnen \& Binney (1998)
by inverting the Poisson equation with the Bessel integrals
(Quinn \& Goodman 1986). 
$\lambda$ is an approximate expression  of  the vertical tilt of the velocity ellipsoid close 
to the Galactic  plane (at z=0).  $\lambda$ is 1  in the case of a spherical potential, when the ellipsoid is pointing towards the Galactic center , while  $\lambda$ is 0 for a cylindrical potential when the ellipsoid
is always parallel to the Galactic  plane.
 Amendt \& Cutterford (1991) show that $\lambda$ 
is related to the mass gradient in  the Galactic plane.
 No vertex deviation is included in the model although a deviation 
decreasing from 25$^0$ for young stars to near zero for an old population has been
found by various authors (Dehnen \& Binney 1998, Bienayme 1999, Soubiran et al 2002). The thick disk is  isothermal with
 $(\sigma_{RR}, \sigma_{\Phi \Phi}, \sigma_{zz})=(70,50,45)$ Km s$^{-1}$  as starting values. 
The $V_{lag}$ is assumed to have a canonical value of 35  Km s$^{-1}$. The 
possibility of simulating a vertical gradient in the rotational velocity
of the thick disk is included as suggested by Chiba \& Beers (2000). 
The halo  has $(\sigma_{RR}, \sigma_{\Phi \Phi}, \sigma_{zz})=(130,95,95)$ Km s$^{-1}$.
The projection matrixes of the space velocities are derived from Mendez et al (2000).
In the following Sections we show a few examples of applications to the GAIA
science. In particular the thin and thick disk are discussed.

\section{Disentangling various stellar populations}

Spectrograph aboard on GAIA operating in the near-IR will measure radial
velocities with an accuracy better
than 2 Km s$^{-1}$ if an high dispersion of 0.25 A$\fdg/$pix is chosen for stars
brighter than V= 16 mag. The expected accuracy will still be better than 8 Km s$^{-1}$ 
for magnitudes as faint as V=17, but will drop to 30  Km s$^{-1}$ 
at V=18 (Zwitter 2002) for a G2V stars. 
This amount of unprecedented good quality data will allow to disentangle thin disk, thick disk
and halo populations, deriving ages, metal content and kinematics.
Fig. 1 presents a simulation of the expected radial velocities  at (l,b)=(270,-45) for
a 2.5 $\times $2.5 deg$^2$ field for stars brighter than V=17, 17.5, and in the magnitude range 17.5-18
at an heliocentric distance of 2000-4000 pc where the contribution of the thick
disk is relevant.
 Expected observational uncertainties
 are included. The thick disk population begins to contribute significantly
 at magnitudes fainter than V=17. In order to disentangle thin and thick disk populations, 
a fainter magnitude limit would be more effective, unless  
kinematic information is coupled with  chemical abundances determinations.

\section {The thick disk velocity gradient}

The presence of a gradient in the thick disk  velocity dispersion
or a multi-component structure is indicative of the formation process
as described in Section 1. From the observational point of
view the situation is far from being clear.
 Soubiran et al (2002) find a thick
disk with a moderate rotational and vertical kinematics, but no vertical gradient suggesting a quick heating of the precursor thin disk. 
Chiba \& Beers (2000) suggest  that far from the plane the thick disk has  lower rotational velocity and higher velocity dispersion than  close
to the Galactic plane. These data are interpreted as 
a vertical gradient of about 30 Km/s/Kpc. Gilmore et al (2002)
propose a different interpretation of the data:  the thick disk has a composite form, indicating the presence
of  relics of disrupted satellites.
Fig. 2 shows a simulation of the thick disk population in a column of 0.09$\times$ 0.09 deg$^2$ at (l,b)=(270,-45) for stars brighter than V=17.5.
Even including the expected GAIA accuracy for an intermediate dispersion of 0.5 A$\fdg/$pix,
still the effect of a vertical gradient of 10 Km/s/Kpc is visible.

\section{The vertical tilt of the thin disk velocity ellipsoid }

Up to now the vertical tilt parameter $\lambda(R)$ 
 of the thin disk velocity ellipsoids 
 is ill-determined.
$\lambda$ is in fact strongly related to the coupling of the U and W velocities
 and  is better constrained by 3-dimensional velocities.
It is found to vary from 0.4 to 0.6 at the solar circle (Amendt \& Cutterford 1991, Bienaym\'e 1999).
Fig.3 shows the expected proper motions and radial velocities
 of the thin disk population under different assumptions for 
$\lambda(R)$, namely 0, 1 and  following Amendt \& Cutterford (1991)
in the direction (l,b)=(26,6). The simulations take into account the  
uncertainties on GAIA determinations for stars brighter than V=17. 
In comparison to the models having  $\lambda(R)=0$ or 1, 
the model with $\lambda(R)$ related to the potential predicts
differences of the mean value of the radial velocities  going from 5 to 15 Km/s, 
depending on the distance.  On the proper motion $\mu_l$ the effect is
at maximum 0.2-0.4 mas/yr, while it  is definitely less than 0.08 mas/yr on $\mu_b$.
 GAIA expected precision on radial velocities and proper motions will be able
to put strong constraints on the determination of the tilt of the velocity ellipsoid.

\begin{figure}
\plotfiddle {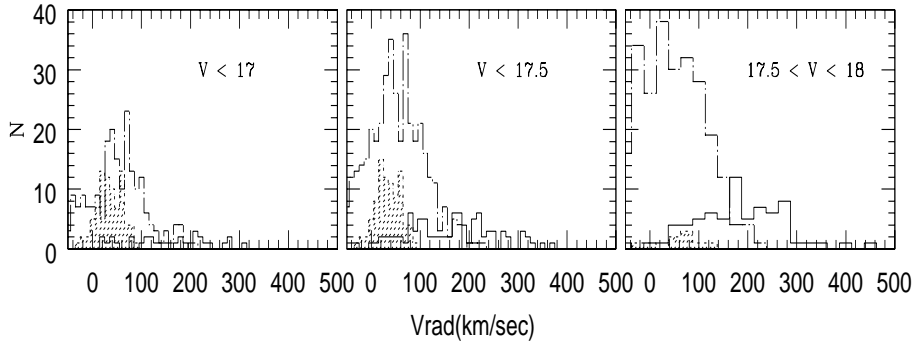} {5truecm}{0} {60}{80}{-190}{-270}
\caption{Simulation of a field at (l,b)=(270,-45) including thin disk (shaded histogram), thick disk
(dashed-dotted line) and halo (solid line) at an heliocentric distance d=2000-4000 pc in various magnitude ranges. The bin size is related to the expected accuracy.}
\end{figure}

\begin{figure}
\plotfiddle {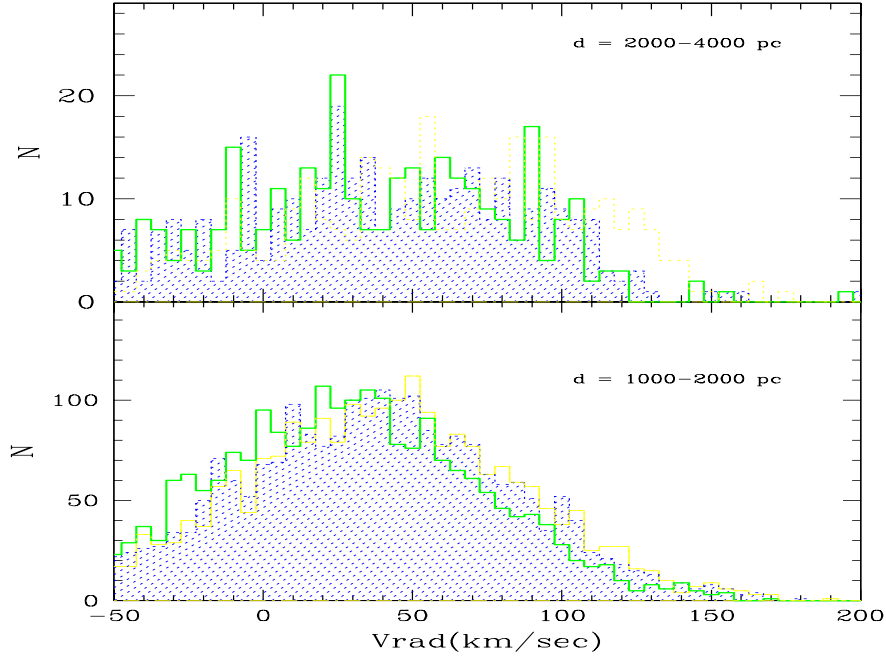} {10truecm}{0}{80}{50}{-250}{-70} 
\caption{
The effect of the thick disk velocity gradient. Dashed line shows a model with
no vertical gradient; heavy solid line presents 
a vertical gradient of 10 Km/s/Kpc, thin solid line is the analogous for 
a vertical gradient of 30 Km/s/Kpc.
d is the heliocentric distance }
\end{figure}

\begin{figure}
\plottwo{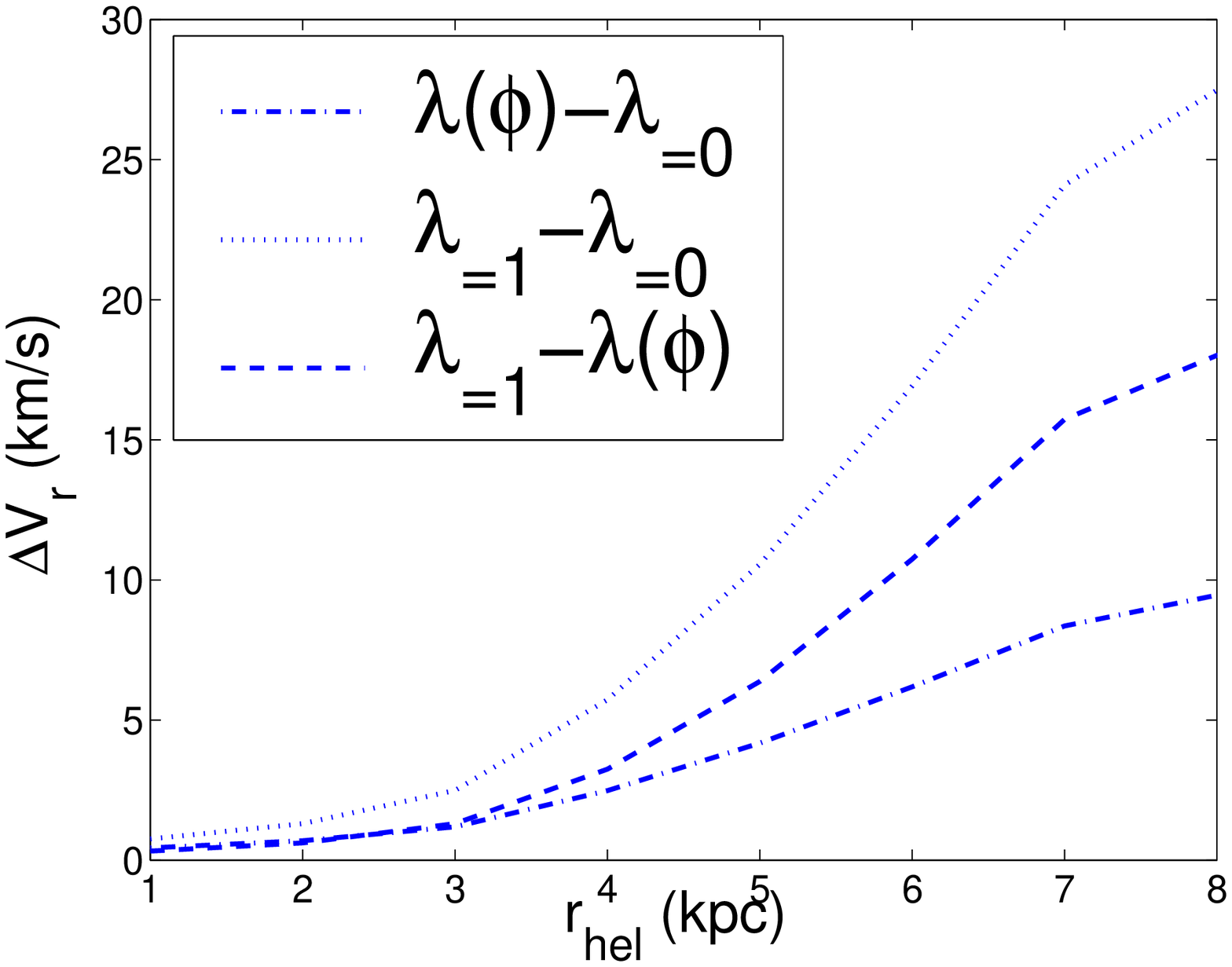}{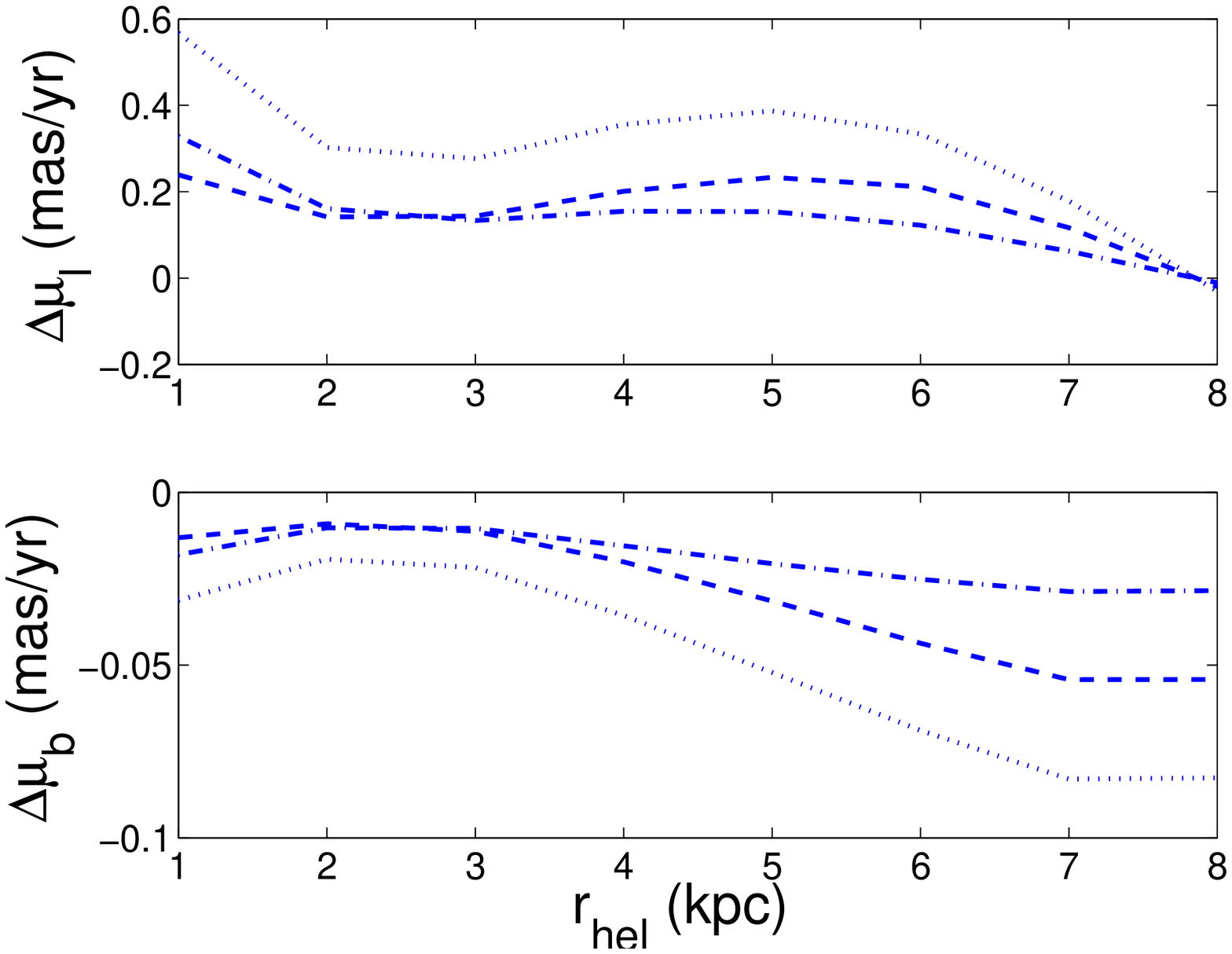}
\caption{
The differences of the mean values  of  V$_{rad}$ and $\mu$ 
at varying $\lambda(R)$
 are plotted as 
functions of the Galactic radius R for magnitudes brighter than V=17. 
 Dotted lines show the difference between the models using $\lambda=0$ and 1;
the dashed-dotted lines are the analogous for  $\lambda=\lambda(R,\Phi)$ and 0; and finally the 
dashed-dotted lines are the analogous for $\lambda=1$,$\lambda(R,\Phi)$.    }
\end{figure}

\section {Conclusions}
  
GAIA data will create a precise 3-dimensional map of  the Galaxy, providing positional information, radial
velocities, luminosity, temperature and chemical composition of
 a representative
sample of stars.
Here
we present a new implementation of the Padova Galaxy Model,
where kinematics simulations are included.
 This model includes a description of the vertical tilt of the ellipsoids
of the velocity of the thin disk following Amendt \& Cutterford (1991),
 in addition to  isothermal thick disk and halo. This formulation might be
specially useful to simulate the kinematics of the thin disk till a
vertical height of 1 Kpc.
A few examples of application to
 the GAIA science are presented, in particular concerning radial velocities
determinations.
The main conclusions are: 

1) the expected accuracy on  radial velocities
will put strong constraints on the determination of the vertical tilt
of the velocity ellipsoids of the thin disk;

2) since the  thick disk begins to contribute significantly
 at magnitudes fainter than V=17, 
to disentangle thin and thick disk populations at high latitude,
a fainter magnitude limit would be more effective, unless  
kinematic information are coupled with  chemical abundances determinations;

3) finally, the expected accuracy on radial velocities 
at intermediate dispersion will allow the determination of a possible 
vertical velocity gradient in the thick disk of
at least 10 Km s$^{-1}$.

\acknowledgements{The authors wish  to thank U. Munari and the Organizing Committee
for the excellent organization of  this extremely stimulating meeting.}

\begin {references}

 \reference Amendt, P., Cutterford, P., 1991, \apj 368, 79

 \reference  Bertelli, G., Bressan, A., Chiosi, C., Ng, Y.K., Ortolani, S., 1995, A\&A, 301, 381 

 \reference Baraffe, I., Chabrier, G., Allard F., Hauschildt P. H., 1998, \aap 337, 403

 \reference Binney J.J., Tremaine, S., 1994, Galactic Dynamics, (Princeton: Princeton University Press) p. 190

 \reference Bienaym\'e, O., 1999, \aap, 341, 86 

 \reference Chiba, M., \& Beers T.C.,  2000, \aj, 119, 2843

 \reference Dehnen , W., \&  Binney, J.J.,  1998, \mnras 298, 387

 \reference Drimmel R,  \& Spergel D.N. 2001, ApJ 556,181

 \reference Freeman, K., \& Hawthorn , J. B.,2002, \araa, 40, 487

\reference Fuchs B.,  \&  Wielen R., 1987, in The Galaxy, NATO-ASI Ser., ed G. Gilmore, B Carwell, Reidel, p375 

\reference Gilmore, G., Wyse R.F.G., \& Norris J.E., 2002, ApJL 574 

 \reference Girardi,L.,  Bressan, A., Bertelli, G., \&  Chiosi C., 2000, A\&AS 141, 371

 \reference Lewis, J. R., \& Freeman, K. C., 1989, \aj 97, 139

 \reference Marigo, P., 2002 \aap  378, 507

 \reference Mendez, R.A., \& van Altena, W. F., 2000, \aj 120, 1161

 \reference Quinn, P.J.,\&  Goodman J., 1986 \apj 309, 472

 \reference Soubiran, C., Bienaym\'e O.,\& Siebert A., 2002, \aap in press

 \reference Steinmetz M., \& Navarro  J.F., 1999, ApJ 513, 555

 \reference Vallenari A., Bertelli G.,\&  Schmidtobreick L., 2000, A\&A 361,73
 
  \reference Zwitter, T. 2002, this conference 
\end{references}

\end{document}